\documentclass{appolb}
\usepackage{graphicx}
\usepackage{braket}

\usepackage{cite}

\newcommand{\Pom}{\mathbb{P}}

\newcommand{\bk}{\mbox{\boldmath $k$}}

\usepackage{amsmath,amsfonts,amssymb,amstext,mathrsfs}

\begin{document}
\title{Exclusive Bremsstrahlung of One and Two Photons in Proton-Proton Collisions
\thanks{Presented by A. Szczurek at XXX Cracow EPIPHANY Conference
on Precision Physics at High Energy Colliders,
Krak{\'o}w, Poland,
January 8-12, 2024.}%
}
\author{Piotr Lebiedowicz, Antoni Szczurek 
\address{Institute of Nuclear Physics Polish Academy of Sciences, 
Radzikowskiego 152, PL-31342 Krak{\'o}w, Poland}
\\[3mm]
{Otto Nachtmann 
\address{Institut f\"ur Theoretische Physik, Universit\"at Heidelberg,
Philosophenweg 16, D-69120 Heidelberg, Germany}
}
\\[3mm]
}
\maketitle
\begin{abstract}
We discuss the diffractive bremsstrahlung of a single photon 
in the $pp \to pp \gamma$ reaction 
at LHC energies and at forward photon rapidities.
We compare the results for our standard approach,
based on QFT and the tensor-Pomeron model,
with two versions of soft-photon approximations, SPA1 and SPA2, where the radiative amplitudes
contain only the leading terms proportional to $\omega^{-1}$
(the inverse of the photon energy).
SPA1, which does not have the correct energy-momentum relations, 
performs surprisingly well in the kinematic range considered,
namely at very forward photon rapidities 
and $0.02 < \xi_{1,2} < 0.1$, the relative energy loss 
of the protons, corresponding to small values of the photon transverse momentum.
Azimuthal correlations between outgoing particles are presented.
We discuss also the role of the $p p \to p p \pi^0$ background for
single photon production.
We discuss also the possibility of a measurement 
of the $pp \to pp \gamma \gamma$ reaction.
Our predictions can be verified by ATLAS-LHCf combined experiments.
\end{abstract}
  
\section{Introduction}
\label{sec:introduction}

In this contribution we will be concerned with
exclusive diffractive photon(s) bremsstrahlung in proton-proton collisions at high energies.
The presentation is based on 
\cite{Lebiedowicz:2022nnn,Lebiedowicz:2023rgc}
where all details and many more results can be found.

The forward photon production
(inclusive) cross-section 
in proton-proton collisions was measured 
with the RHICf detector \cite{Adriani:2022csq}
at centre-of-mass energy $\sqrt{s} = 510$~GeV 
and with the LHCf detector at $\sqrt{s} = 0.9, 7$ 
and 13~TeV \cite{LHCf:2017fnw}.
The LHCf experiment is designed to measure the production
cross section of neutral particles
in the pseudorapidity region $|\eta| > 8.4$, up to zero-degree.
Several joint analyses with the ATLAS-LHCf detectors
are planned; 
see \cite{Tiberio:2023xgj} and references therein.
In contrast, the exclusive reaction $pp \to pp \gamma$ 
has not yet been identified experimentally.
Some feasibility studies 
for the measurement of the exclusive bremsstrahlung cross-sections
were performed for RHIC energies 
\cite{Chwastowski:2015mua} and for LHC energies 
using the ATLAS forward detectors
\cite{Chwastowski:2016jkl,Chwastowski:2016zzl}.

The theoretical methods which we use
in our analysis were developed
by us in \cite{Lebiedowicz:2021byo,Lebiedowicz:2022nnn}.
In \cite{Lebiedowicz:2021byo} 
we discussed the soft-photon radiation in pion-pion scattering.
There we compared our standard (called also ``exact'') model results for diffractive photon-bremsstrahlung 
to various soft-photon approximations (SPAs) based on
the soft-photon theorems.
In \cite{Lebiedowicz:2022nnn} we extended these considerations
to the $pp \to pp \gamma$ reaction at $\sqrt{s} = 13$~TeV.
In contrast to the analysis of this reaction 
previously discussed by two of us \cite{Lebiedowicz:2013xlb} 
in our recent works we use the tensor-Pomeron model developed in \cite{Ewerz:2013kda,Ewerz:2016onn}.
The tensor-Pomeron model was successfully applied to many hadronic reactions, in particular to central exclusive diffractive production processes;
see e.g. \cite{Lebiedowicz:2016ioh}.
From \cite{Ewerz:2013kda,Ewerz:2016onn} 
we know the form of the effective Pomeron propagator
and the Pomeron-proton-proton vertex function.
We have shown in \cite{Lebiedowicz:2023mhe}
that the photon-Pomeron fusion mechanism
does not play an important role
at very forward photon rapidities, 
which is what we are most interested in.
In our analysis of photon-bremsstrahlung processes 
\cite{Lebiedowicz:2021byo,Lebiedowicz:2022nnn}
we respect the general QFT structure of the radiative amplitudes.

It is worth noting that for $\omega \to 0$,
in the region where soft-photon theorems discussed in\cite{Lebiedowicz:2023ell} 
should be applicable,
our bremsstrahlung distributions presented below and in 
\cite{Lebiedowicz:2022nnn,Lebiedowicz:2023rgc}
are an exact result of QCD plus lowest order electromagnetism.

The theoretical methods which we developed for
the exclusive diffractive bremsstrahlung of soft photons
can be also used for the production of ``dark photons''.
For some interesting scenarios in this context, 
we refer the reader to 
Refs.~\cite{Foroughi-Abari:2021zbm,Shin:2021bvz,Gorbunov:2023jnx}.
Searching for a signal of light, 
weakly-interacting dark photons
and many other long-lived particles of new physics,
in the very forward region,
is the main task of the FASER detector at the LHC 
\cite{FASER:2023tle}.

\section{Sketch of the formalism}
\label{sec:formalism}

\subsection{$pp \to pp \gamma$}
\label{sec:2a}

We consider the reaction
\begin{eqnarray}
p (p_{a},\lambda_{a}) + p (p_{b},\lambda_{b}) \to 
p (p_{1}',\lambda_{1}) + p (p_{2}',\lambda_{2}) + \gamma(k, \epsilon)
\label{pp_ppgam}
\end{eqnarray}
at high energies and small momentum transfers.
The momenta are indicated in brackets,
the helicities of the protons are denoted by
$\lambda_{a}, \lambda_{b}, \lambda_{1}, \lambda_{2} 
\in \{1/2, -1/2 \}$,
and $\epsilon$ is the polarization vector of the photon.
The kinematic variables are
\begin{eqnarray}
&& s = (p_{a} + p_{b})^{2} = (p_{1}' + p_{2}' + k)^{2}\,, \nonumber \\
&& t_{1} = (p_{a} - p_{1}')^{2} \,, 
\quad t_{2} = (p_{b} - p_{2}')^{2} \,.
\label{2.17}
\end{eqnarray}
The rapidity of the photon is then
\begin{eqnarray}
{\rm y} = \frac{1}{2} \ln \frac{k^{0} + k^{3}}{k^{0} - k^{3}} 
= - \ln \tan \frac{\theta}{2} \; ,
\label{rapidity}
\end{eqnarray}
where $\theta$ is the polar angle of $\bk$, 
$\cos\theta = k^{3}/|\bk|$.
The proton relative energy-loss parameters
can be expressed by the kinematical variables of the photon,
\begin{eqnarray}
\xi_{1} = \frac{k_{\perp}}{\sqrt{s}} \exp({\rm y})
+ {\cal O}\left(\frac{M^{2}}{s} \right)\,, \quad
\xi_{2} = \frac{k_{\perp}}{\sqrt{s}} \exp(-{\rm y})
+ {\cal O}\left(\frac{M^{2}}{s} \right)\,.
\label{2.19}
\end{eqnarray}
%

\begin{figure}[htb]
($a$)\includegraphics[width=0.24\textwidth]{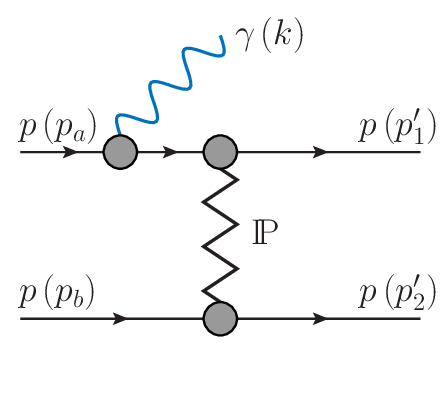}
($b$)\includegraphics[width=0.28\textwidth]{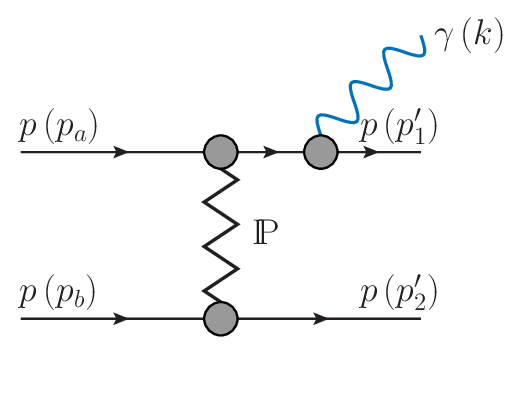}
($c$)\includegraphics[width=0.24\textwidth]{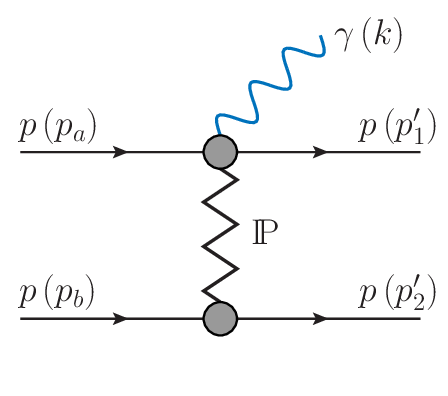}
\caption{Diagrams for the reaction $pp \to pp \gamma$
via diffractive bremsstrahlung
with exchange of the Pomeron $\Pom$.
In addition there are diagrams $(d)$, $(e)$, $(f)$
corresponding to emission of photon from the $p_{b}$-$p_{2}'$ line.
These are not shown here.}
\label{fig:1}
\end{figure}

The cross section for the photon yield can be calculated as follows
\begin{eqnarray}
&&d\sigma({pp \to pp \gamma}) =
\frac{1}{2\sqrt{s(s-4 m_{p}^{2})}}
\frac{d^{3}k}{(2 \pi)^{3} \,2 k^{0}}
\int 
\frac{d^{3}p_{1}'}{(2 \pi)^{3} \,2 p_{1}'^{0}}
\frac{d^{3}p_{2}'}{(2 \pi)^{3} \,2 p_{2}'^{0}}
\nonumber \\
&& \qquad  \times 
(2 \pi)^{4} \delta^{(4)}(p_{1}'+p_{2}'+k-p_{a}-p_{b})
\frac{1}{4}
\sum_{p\; {\rm spins}}
{\cal M}_{\mu} {\cal M}_{\nu}^{*} (-g^{\mu \nu})\,;  \qquad
\label{xs_2to3}
\end{eqnarray}
see Eqs.~(2.33)--(2.35) of \cite{Lebiedowicz:2022nnn}.
Above $k^{0} \equiv \omega$,
${\cal M}_{\mu}$ is the radiative amplitude.
In the following, we consider only the leading Pomeron-exchange contribution
within the tensor-Pomeron approach \cite{Ewerz:2013kda,Ewerz:2016onn}.
Our diffractive photon-bremsstrahlung amplitude
includes 6 diagrams (see Fig.~1)
${\cal M}_{\mu} = {\cal M}_{\mu}^{(a)}
+ {\cal M}_{\mu}^{(b)} + {\cal M}_{\mu}^{(c)}
+ {\cal M}_{\mu}^{(d)} + {\cal M}_{\mu}^{(e)}
+ {\cal M}_{\mu}^{(f)}$;
see (2.62) and (B3) of \cite{Lebiedowicz:2022nnn}.
The amplitudes $(a, b, d, e)$ corresponding to photon emission 
from the external protons are determined 
by the off-shell $pp$ elastic scattering amplitude. 
The contact terms $(c, f)$ are needed 
in order to satisfy gauge-invariance constraints.
With the off-shell $pp$ elastic scattering amplitude 
${\cal M}^{(0)}$,
the standard proton propagator $S_{F}$,
and the $\gamma pp$ vertex function,
we get the following radiative amplitudes:
\begin{eqnarray}
{\cal M}_{\mu}^{(a)}&=&
-
\bar{u}_{1'} \otimes \bar{u}_{2'} \,
{\cal M}^{(0)}(p_{a}-k,p_{b},p_{1}',p_{2}')\,
\big(
S_{F}(p_{a}-k)
\Gamma_{\mu}^{(\gamma pp)}(p_{a}-k,p_{a})\,
u_{a} \big) \otimes u_{b}\,, 
\label{2.26} \nonumber \\
{\cal M}_{\mu}^{(b)} &=&
-
\big( \bar{u}_{1'} 
\Gamma_{\mu}^{(\gamma pp)}(p_{1}',p_{1}'+k)
S_{F}(p_{1}'+k) \big)
\otimes \bar{u}_{2'} \,
{\cal M}^{(0)}(p_{a},p_{b},p_{1}'+k,p_{2}')\,
u_{a} \otimes u_{b}\,. \nonumber \\ 
\label{2.27}
\end{eqnarray}
In our tensor-product notation,
the first factors will always refer to the \mbox{$p_{a}$-$p_{1}'$} line,
and the second refer to the $p_{b}$-$p_{2}'$ line.
Using the Ward-Takahashi identity 
\begin{equation}
(p'-p)^{\mu} \Gamma_{\mu}^{(\gamma pp)}(p',p) =
-e \big[ S_{F}^{-1}(p') - S_{F}^{-1}(p) \big]\,,
\label{2.25}
\end{equation}
and imposing the gauge-invariance condition
$k^{\mu} ({\cal M}_{\mu}^{(a)}
+ {\cal M}_{\mu}^{(b)} + {\cal M}_{\mu}^{(c)}) = 0$
we obtain
\begin{eqnarray}
k^{\mu} {\cal M}_{\mu}^{(c)} = 
e \bar{u}_{1'} \otimes \bar{u}_{2'}
\big[ {\cal M}^{(0)}(p_{a}-k,p_{b},p_{1}',p_{2}') 
     - {\cal M}^{(0)}(p_{a},p_{b},p_{1}'+k,p_{2}')\big] u_{a} \otimes u_{b}\,. \quad 
\label{2.31}
\end{eqnarray}
In a similar way we proceed with the amplitudes $(d)$, $(e)$, and $(f)$.
For details how to calculate above results
we refer the reader to Sec.~II~C 
and Appendix~B of \cite{Lebiedowicz:2022nnn}.

We compare our standard results
to two soft-photon approximations, SPA1 and SPA2,
where we keep only the pole terms $\propto \omega^{-1}$.
In SPA1, the radiative amplitude has the form
\begin{eqnarray}
{\cal M}_{\mu, \;{\rm SPA1}}
= e{\cal M}^{({\rm on\; shell})\,pp}(s,t)
\Big[ 
-\frac{p_{a \mu}}{(p_{a} \cdot k)}
+\frac{p_{1 \mu}}{(p_{1} \cdot k)}
-\frac{p_{b \mu}}{(p_{b} \cdot k)}
+\frac{p_{2 \mu}}{(p_{2} \cdot k)} \Big], \quad
\label{SPA1}
\end{eqnarray}
where ${\cal M}^{({\rm on\; shell})\,pp}(s,t)$ 
is the amplitude for on-shell $pp$-scattering.
In the SPA1, the photon momentum $k$ was, on purpose,
omitted in the energy-momentum conserving $\delta$ function
in the evaluation of the cross section.

In the SPA2, the correct $2 \to 3$ kinematics is used,
that is,
we keep the exact energy-momentum relation $p_{a} + p_{b} = p_{1}' + p_{2}' + k$.
We calculate the photon yield using (\ref{xs_2to3}) 
with the radiative amplitude as follows
\begin{eqnarray}
{\cal M}_{\mu, \;{\rm SPA2}}
= {\cal M}_{{\Pom},\mu}^{(a + b + c)\,1}(s,t_{2})
+
{\cal M}_{{\Pom},\mu}^{(d + e + f)\,1}(s,t_{1})\,.
\label{SPA2}
\end{eqnarray}
The explicit expressions of these terms
are given by (3.4), (B4), (B15) of \cite{Lebiedowicz:2022nnn}.

\subsection{$pp \to pp \gamma \gamma$}
\label{sec:2b}

\begin{figure}[!h]
\includegraphics[width=1\textwidth]{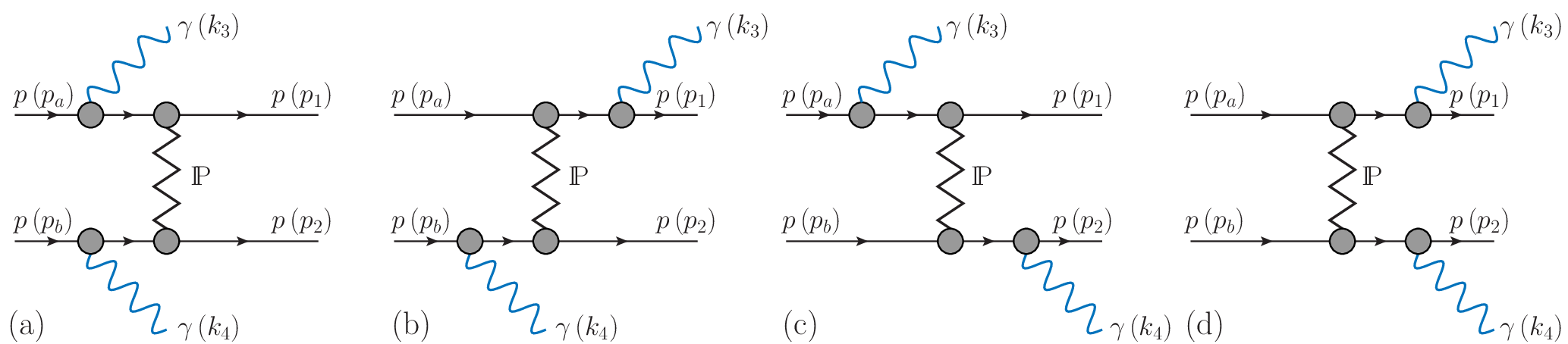}
\caption{Diffractive two-photon bremsstrahlung diagrams 
for the reaction $pp \to pp \gamma \gamma$.
These four diagrams contribute to the SPA1 amplitude
(\ref{SPA1_gamgam}).}
\label{fig:pp_brems_gamgam_SPA1}
\end{figure}

Here we consider two-photon bremsstrahlung in the reaction 
(see Fig.~\ref{fig:pp_brems_gamgam_SPA1})
\begin{eqnarray}
p (p_{a}) + p (p_{b}) \to 
p (p_{1}') + p (p_{2}') + \gamma(k_{3})+ \gamma(k_{4})\,.
\label{pp_ppgamgam}
\end{eqnarray}
%
For the calculation of amplitude (\ref{pp_ppgamgam}) 
we use SPA1:
\begin{eqnarray}
{\cal M}_{\mu \nu, \;{\rm SPA1}}
&=& e^{2} {\cal M}^{({\rm on\; shell})\,pp}(s,t) \nonumber \\
&& \times
\Big[ 
-\frac{p_{a \mu}}{(p_{a} \cdot k_{3})}
+\frac{p_{1 \mu}}{(p_{1} \cdot k_{3})} \Big]
\Big[ 
-\frac{p_{b \nu}}{(p_{b} \cdot k_{4})}
+\frac{p_{2 \nu}}{(p_{2} \cdot k_{4})} \Big]\,. \quad
\label{SPA1_gamgam}
\end{eqnarray}
In this case, we shall require that one photon is emitted 
at forward and one at backward rapidities,
\mbox{$8.5 < {\rm y}_{3} < 9$} 
and $-9 < {\rm y}_{4} < -8.5$,
respectively, and that $0.02 < \xi_{1,2} < 0.1$,
where we define
\begin{eqnarray}
&&\xi_{1,2} = \frac{k_{\perp 3}}{\sqrt{s}} \exp(\pm {\rm y}_{3})
        + \frac{k_{\perp 4}}{\sqrt{s}} \exp(\pm {\rm y}_{4})\,.
\label{xi_2to4}
\end{eqnarray}

\section{Results}
\label{sec:results}

In Fig.~\ref{fig:1} we show the distributions for our standard approach for the $pp \to pp \gamma$ reaction 
together with the results obtained 
via SPA1~(\ref{SPA1}) and SPA2~(\ref{SPA2}).
Recall that in both SPAs
we keep only the pole terms $\propto \omega^{-1}$
in the radiative amplitudes.
Bremsstrahlung photons
are emitted predominantly in very forward-rapidity region 
$9 < {\rm y} < 10$
and with small values of photon transverse momentum $k_{\perp}$.
Photons can be measured by the LHCf detectors,
in the region $|{\rm y}| > 8.4$,
and the protons by the ATLAS forward proton spectrometers (AFP).
Due to the limitation $0.02 < \xi_{1} < 0.1$
the energy of the photons is limited to 
$130~{\rm GeV} < \omega < 650~{\rm GeV}$.
The SPA1 result performs surprisingly well in the kinematic range considered.
For the SPA2, the deviations from our exact (standard) result
increase rapidly with growing $\omega$ and $k_{\perp}$.
There is an important the interference 
between the leading term $\propto \omega^{-1}$
and the non-leading terms occurring in the radiative amplitudes.
For more details on the size of various contributions
we refer to the discussions in \cite{Lebiedowicz:2022nnn}
and Fig.~17 therein.

\begin{figure}[htb]
\includegraphics[width=0.45\textwidth]{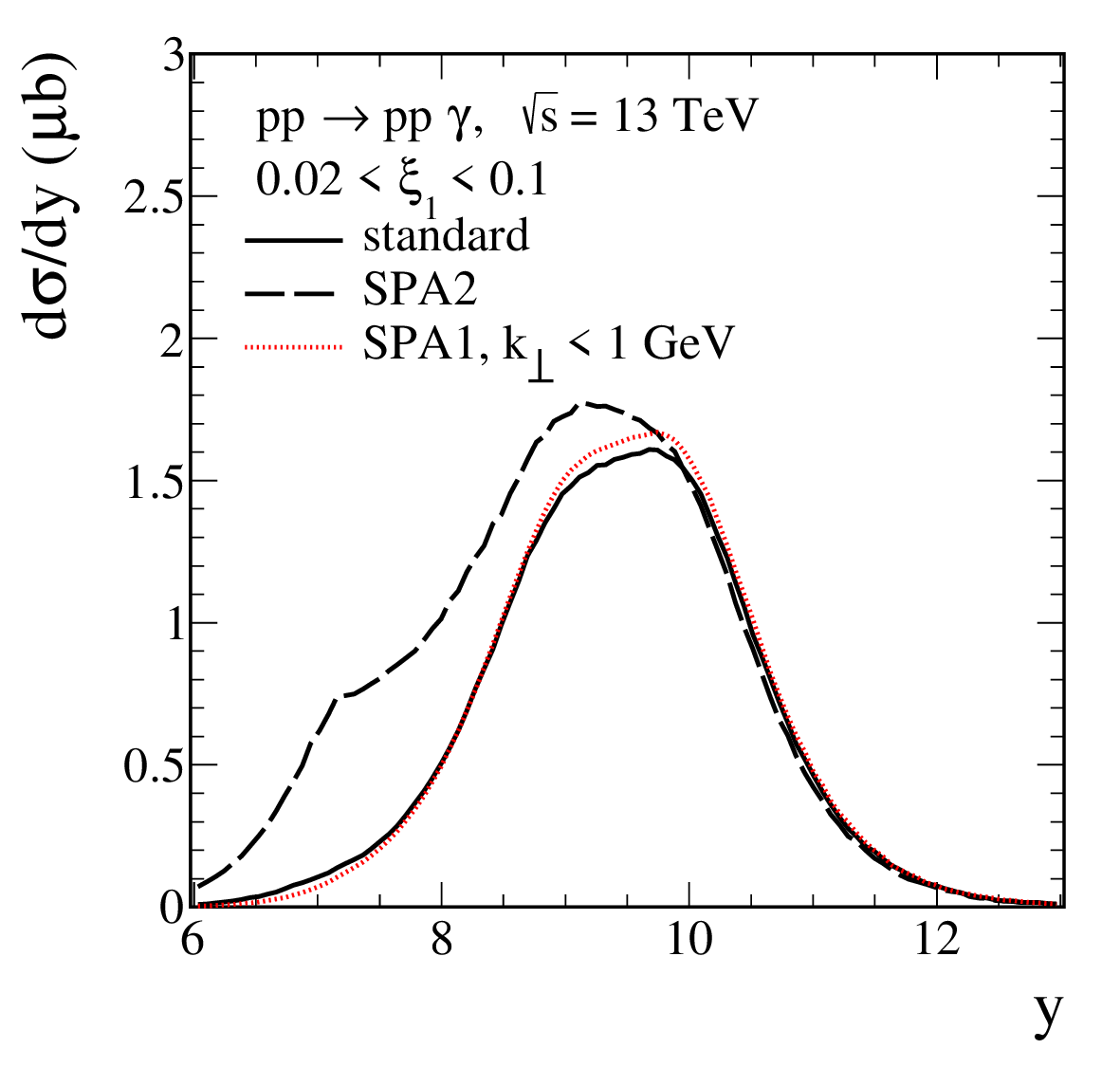}
\includegraphics[width=0.45\textwidth]{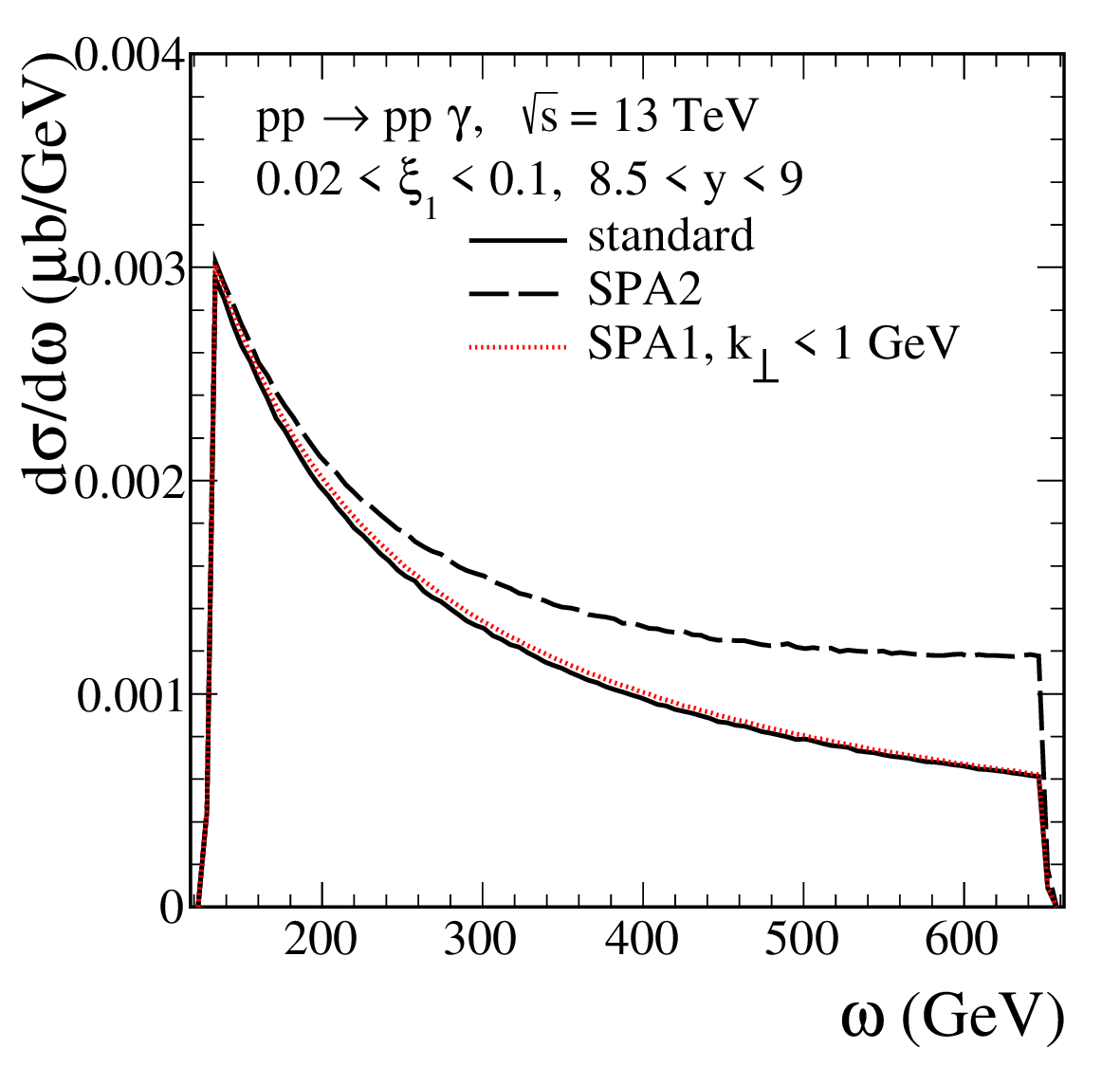}
\caption{The differential distributions 
in the rapidity of the photon (left panel) and 
in the energy of the photon (right panel)
for the $pp \to pp \gamma$ reaction.
The solid line corresponds to complete (standard)
bremsstrahlung model (as in Fig.~\ref{fig:1}),
the black long-dashed line corresponds to SPA2 (see Eq.~(\ref{SPA2})),
and the red dotted line corresponds to SPA1 
(see Eq.~(\ref{SPA1})).}
\label{fig:2}
\end{figure}

\begin{figure}[!ht]
\includegraphics[width=0.45\textwidth]{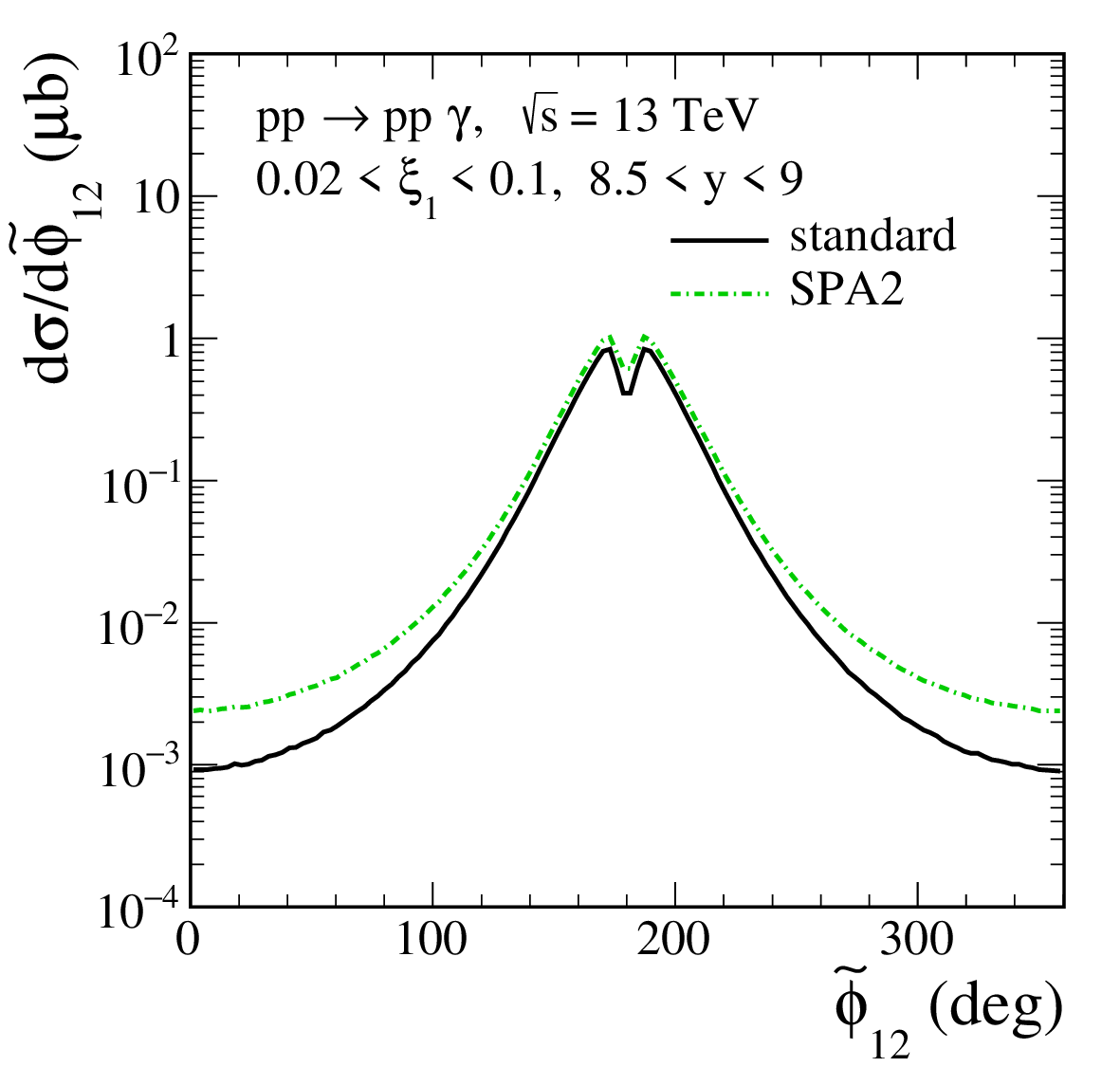}
\includegraphics[width=0.45\textwidth]{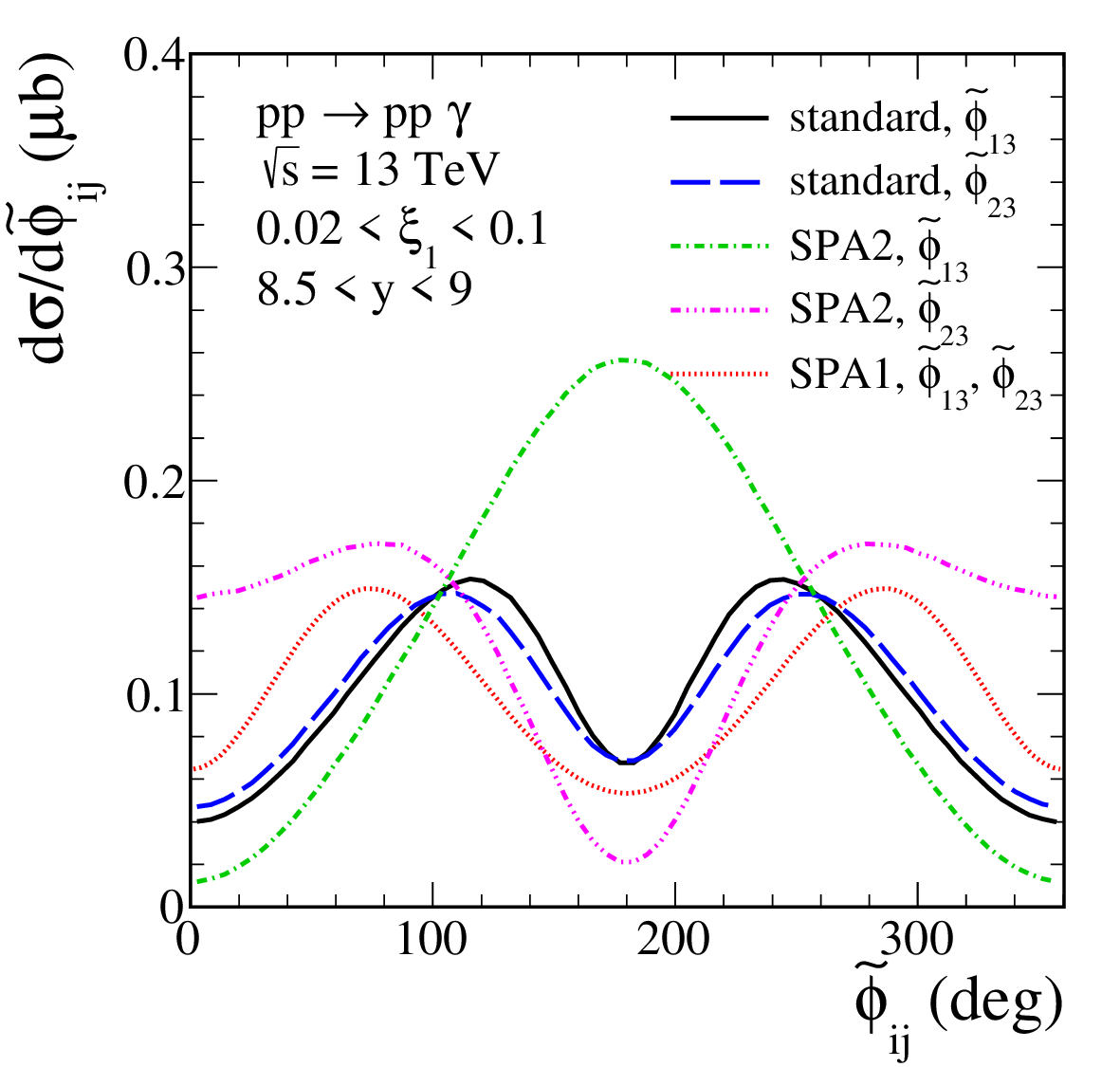}
\caption{\label{fig:3}
\small
The distributions in the angles 
$\tilde{\phi}_{ij}$ defined by (\ref{phi_ij})
for the $pp \to pp \gamma$ reaction.
The results for the standard (exact) bremsstrahlung model 
and the SPAs are shown.}
\end{figure}

The azimuthal correlations between outgoing particles
are particularly interesting.
Figure~\ref{fig:3} shows
the distributions in 
$\tilde{\phi}_{ij}$ defined as 
\begin{eqnarray}
\tilde{\phi}_{ij} = \phi_{i} - \phi_{j} \quad {\rm mod}(2 \pi)\,,
\qquad 0 \leqslant \tilde{\phi}_{ij} < 2 \pi\,.
\label{phi_ij}
\end{eqnarray}
In the left panel we show the results in $\tilde{\phi}_{12}$,
the angle between 
the transverse momenta of the outgoing protons,
for our standard and SPA2 calculations.
For SPA1 (not shown here)
the outgoing protons are back-to-back,
$\tilde{\phi}_{12} = \pi$,
i.e., the outgoing protons and the beam are
in one plane ${\cal S}_{0}$.
For our standard approach the main contribution
is also placed at $\tilde{\phi}_{12} \approx \pi$.
The right panel of Fig.~\ref{fig:3} shows the distributions
in $\tilde{\phi}_{13}$ ($\tilde{\phi}_{23}$),
the azimuthal angles between the proton $p(p_{1}')$ ($p(p_{2}')$)
and the photon $\gamma(k)$.
The SPA1 and our standard results show maxima 
for $\tilde{\phi}_{13}$ and $\tilde{\phi}_{23}$
around $\pi/2$ and $3\pi/2$.
This corresponds to emission of the photon 
in a plane ${\cal S}_{1}$
which is orthogonal to the plane ${\cal S}_{0}$.
The SPA2 results
for the $\tilde{\phi}_{13}$ and $\tilde{\phi}_{23}$
distributions deviate very significantly 
from our standard results.

\begin{figure}[!ht]
\includegraphics[width=0.45\textwidth]{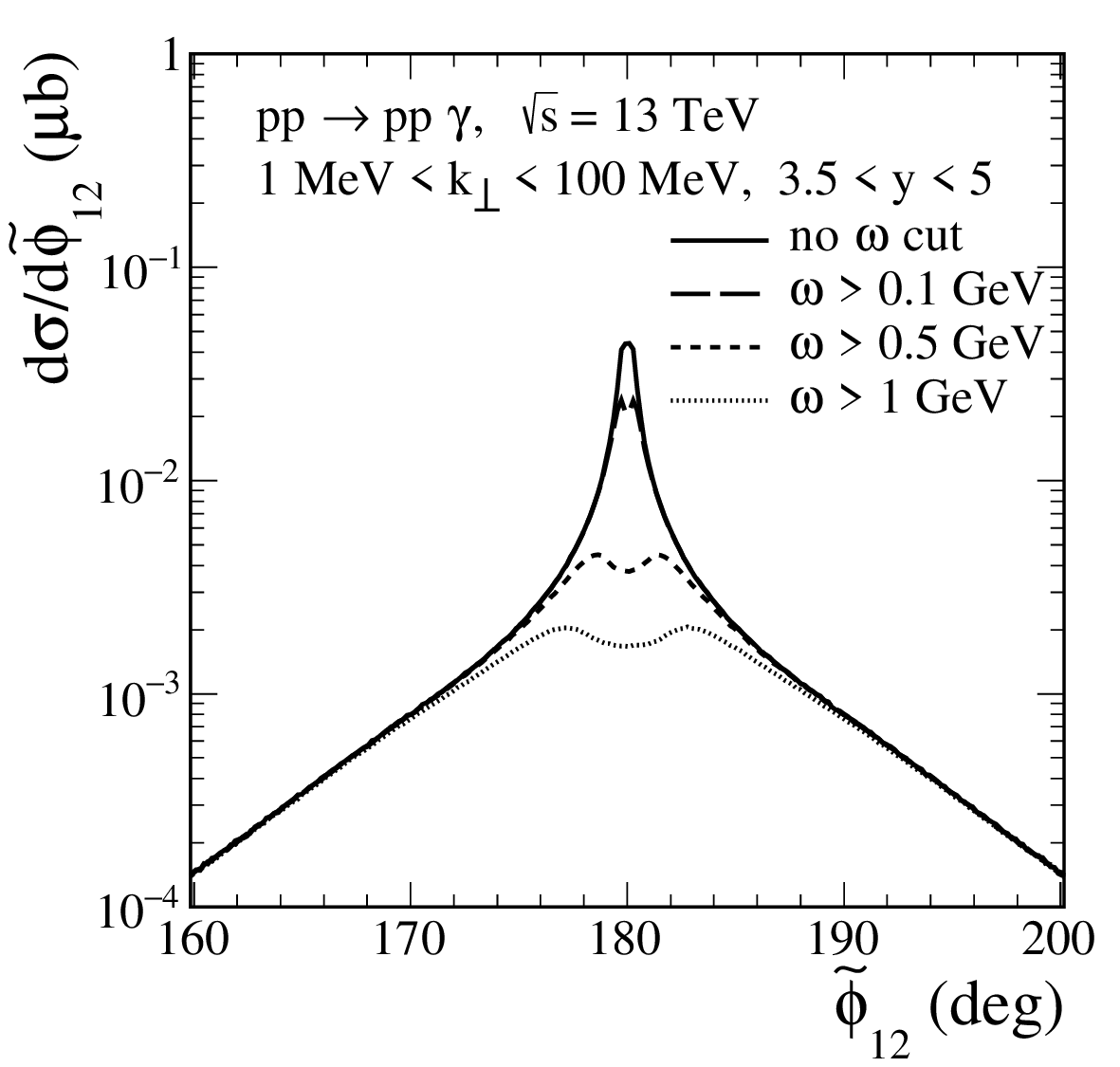}
\includegraphics[width=0.45\textwidth]{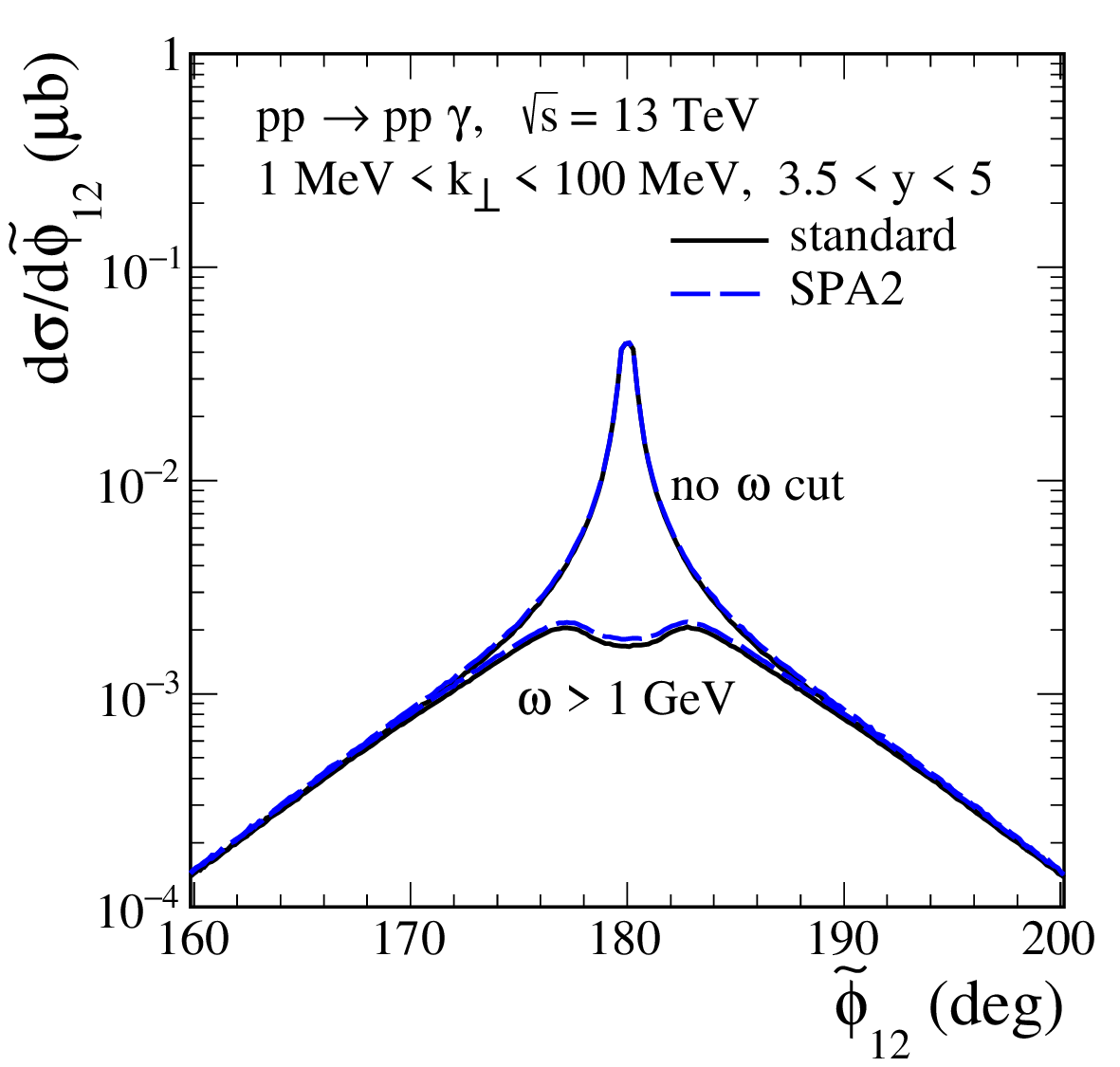}\\
\includegraphics[width=0.45\textwidth]{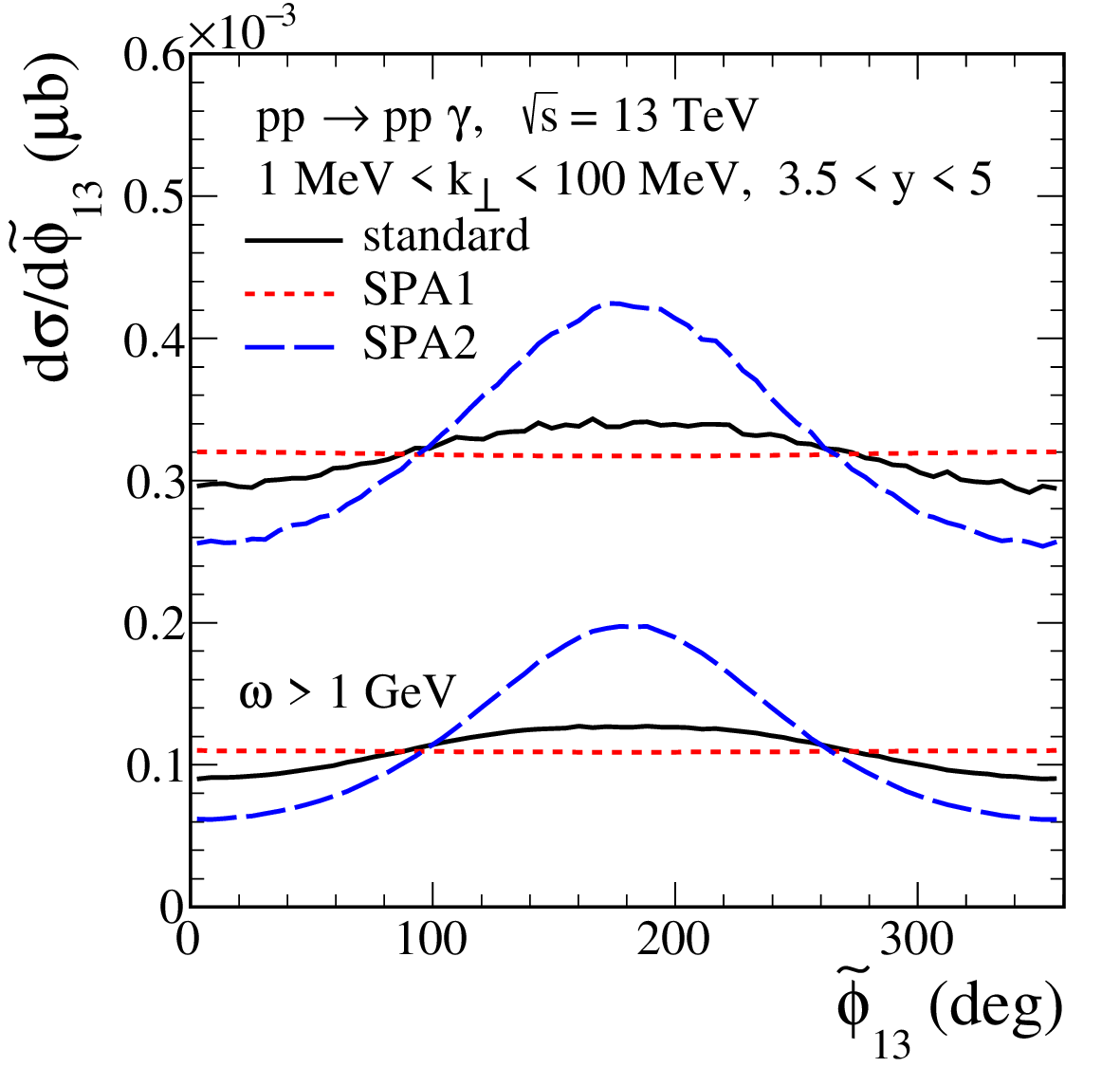}
\includegraphics[width=0.45\textwidth]{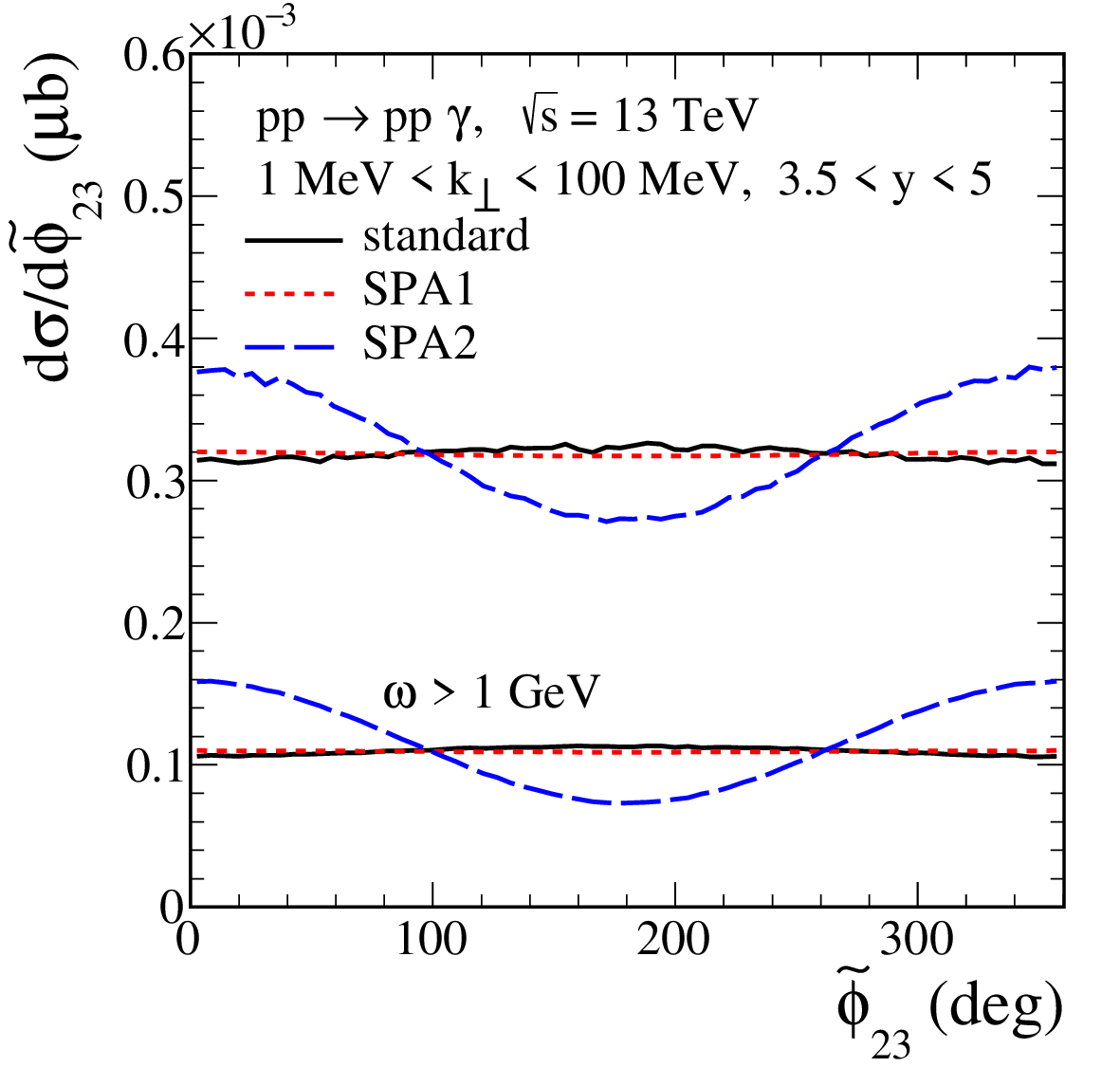}
\caption{\label{fig:ALICE}
\small
The distributions in the angles $\tilde{\phi}_{ij}$
defined by Eq.~(\ref{phi_ij})
for standard bremsstrahlung model (black solid lines),
SPA1 (red dotted lines), and SPA2 (blue dashed lines).}
\end{figure}

Here we will show also first predictions relevant for
future experiment ALICE~3 \cite{ALICE:2022wwr}.
Figure~\ref{fig:ALICE} shows the azimuthal distributions 
$\tilde{\phi}_{ij}$ calculated
for the photon rapidity $3.5 < {\rm y} < 5$,
for the absolute value of the transverse momentum of the photon
$1~{\rm MeV} < k_{\perp} < 100~{\rm MeV}$,
and with various cuts on $\omega$ specified in the figure legends.
The standard results
are compared with those from the approximations, SPA1 and SPA2.
The SPA1 and SPA2 results for 
$\tilde{\phi}_{13}$ and $\tilde{\phi}_{23}$ distributions
deviate from the exact bremsstrahlung results,
and it does not depend significantly on the cut on $\omega$
(see the bottom panels).
For detailed comparisons of the predictions with experiment
and in order to distinguish standard and the approximate results
measurement of the outgoing protons would be welcome.

\begin{figure}[!ht]
\includegraphics[width=0.45\textwidth]{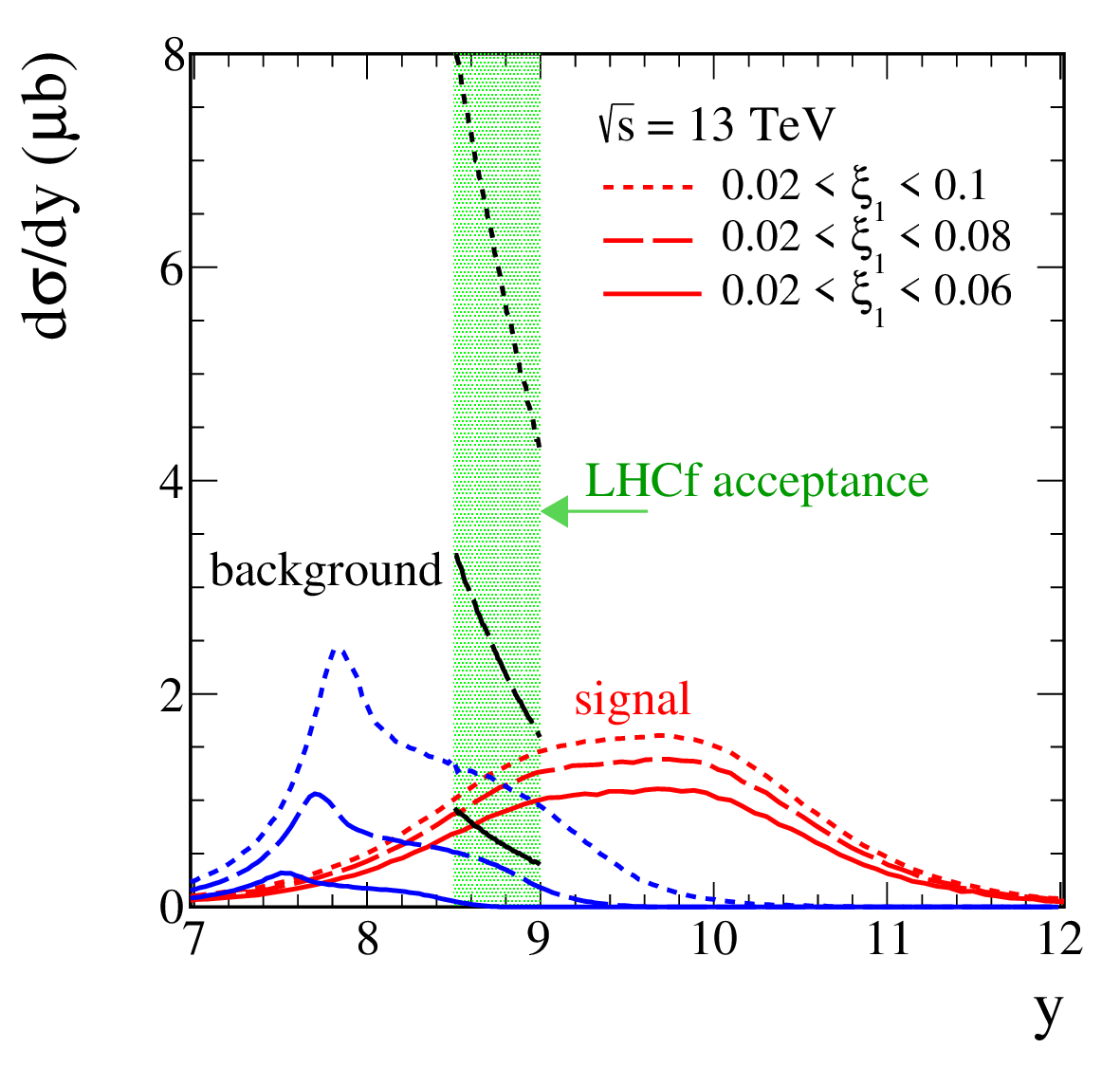}
\includegraphics[width=0.45\textwidth]{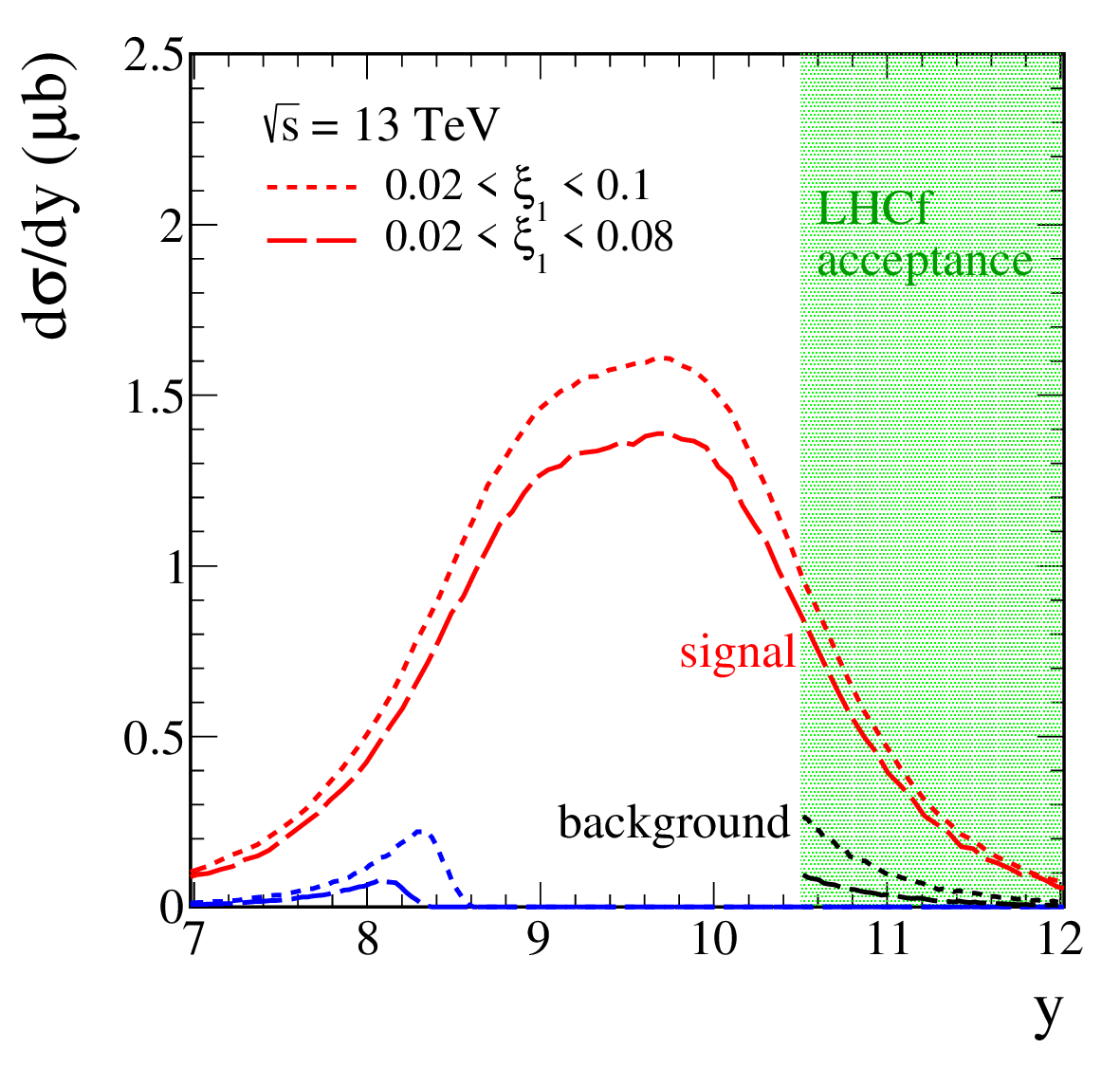}
\caption{
The distribution in rapidity of the photon
for the signal (bremsstrahlung) and background contributions
for different ranges of $\xi_1$. 
The LHCf acceptance regions are marked by the green shaded areas.
In the left panel, the results within the LHCf acceptance
correspond to the acceptance region $8.5 < {\rm y} < 9$,
while for the right panel to ${\rm y} > 10.5$.
For the background contribution
we show the distributions of both photons from the decay of $\pi^0$.
The distributions of the first (measured) photon 
correspond to the black lines,
while the distributions of the second photon
correspond to the blue lines.
Results for the three $\xi_1$ intervals 
as indicated in the figure legends
both for the signal and background contributions are shown.}
\label{fig:background}
\end{figure}

In Fig.~\ref{fig:background} we show the result of a study
of the $p p \to p p (\pi^0 \to \gamma \gamma)$ background
to our reaction $p p \to p p \gamma$. 
We take the upper estimate
of the cross section for the $p p \to p p \pi^0$ reaction
(Drell-Hiida-Deck type model) studied by two of us some
time ago \cite{Lebiedowicz:2013vya}.
In the present calculations, 
we take the background corresponding to the form-factor parameters
$\Lambda_{N} = \Lambda_{\pi} = 1$~GeV 
and without absorption effects.
The red lines represent the distributions of 
our (signal) standard bremsstrahlung of a single photon 
associated with a proton 
for which the fractional energy loss $\xi_1$ 
is in the intervals specified in the figure legend.
In the left panel of Fig.~\ref{fig:background},
for the background contribution we assume that
one photon is measured by the LHCf
in the rapidity interval $8.5 < {\rm y} < 9$, 
$\omega > 130$~GeV, and up to
$\omega_{\rm max} \approx \frac{\sqrt{s}}{2} \, \xi_{1, \rm max}$.
The distributions of the measured photon 
correspond to the black lines (within the LHCf acceptance),
while the distributions of the unmeasured photon
correspond to the blue lines.
The percentage of measured/unmeasured photons depends on 
the upper limit of $\xi_1$. 
For $\xi_{1,\rm max} = 0.06$ (the solid lines) 
the second photon practically cannot be measured.
The signal-to-background ratio for
$\xi_{1,\rm max} = 0.06$ is somewhat larger than 1.
In the right panel of Fig.~\ref{fig:background},
we present the contribution of the background 
for the ${\rm y} > 10.5$ LHCf acceptance range.
Here, the signal-to-background ratio is of order of 4
for $\xi_{1,\rm max} = 0.1$,
and about 10 for $\xi_{1,\rm max} = 0.08$.

We have estimated the coincidence cross section
for the $pp \to pp \gamma \gamma$
reaction only within the SPA1 approach.
We have required that the final state protons and photons
can be measured by the ATLAS forward proton (AFP) spectrometers 
and LHCf detectors, respectively.
We have imposed the kinematical cuts
$8.5 < {\rm y}_{3} < 9$, $-9 < {\rm y}_{4} < -8.5$,
$0.02 < \xi_{1,2} < 0.1$,
and obtained the cross section 
$\sigma \simeq 0.03$~nb for $\sqrt{s} = 13$~TeV.
Hopefully, our predictions will be verified by the ATLAS-LHCf measurement.

\section{Conclusions}

\begin{itemize}

\item
The calculations for the $pp \to pp \gamma$
and $pp \to pp \gamma\gamma$ reactions have been performed
in the tensor-Pomeron model.
We have studied single- and double-photon bremsstrahlung
at very-forward/backward photon rapidities in proton-proton
collisions at the LHC.
We have compared our standard (complete) bremsstrahlung results
and the results using the approximations SPA1 and SPA2.

\item
We have studied the azimuthal angle correlations between outgoing particles. We observe very interesting correlations between protons and photons. Detailed comparisons of our predictions
with experiments (ATLAS-LHCf, ALICE 3) in order to distinguish
our complete and the approximate results measurement of the outgoing protons would be most welcome.

\item
Experimental studies of one photon in $pp$ collisions
available at the ATLAS-LHCf measurement
should provide new information about 
the exclusive diffractive contribution.
We have briefly estimated the background contribution
due to the $pp \to pp \pi^{0}$ diffractive process
for single photon bremsstrahlung.
We have compared the signal and background contributions
in two LHCf acceptance regions.
We conclude that there is a chance to measure single-photon
bremsstrahlung.

\item
The single photon bremsstrahlung mechanism should be identifiable
by the measurement of proton and photon
on one side 
by the ATLAS forward proton spectrometers (AFP) and LHCf detectors, respectively,
and by checking the exclusivity condition 
(no particles in the main detector)
without explicit measurement of the opposite side proton by AFP.
Whether this is sufficient requires further studies,
since such a measurement will probably include one-side diffractive
dissociation, which can be of the order of 25\%.
The bremsstrahlung cross-section for two photons on different sides is rather small
but should be measurable. 

\end{itemize}



\begin{thebibliography}{99}

\bibitem{Lebiedowicz:2023rgc}
P. Lebiedowicz, O. Nachtmann, A. Szczurek,
\textit{Phys. Lett. B} {\bfseries 843}, 138053 (2023).

\bibitem{Lebiedowicz:2022nnn}
P. Lebiedowicz, O. Nachtmann, A. Szczurek,
\textit{Phys. Rev. D} {\bfseries 106}, 034023 (2022).

\bibitem{Chwastowski:2015mua}
J. Chwastowski, A. Cyz, {\L}. Fulek, R. Kycia, B. Pawlik, R. Sikora, J. Turnau, \textit{Acta Phys. Pol. B} {\bfseries 46}, 1979 (2015).

\bibitem{Chwastowski:2016jkl}
J. J. Chwastowski, S. Czekierda, R. Kycia, R. Staszewski, J. Turnau, M. Trzebi{\'n}ski, \textit{Eur. Phys. J. C} {\bfseries 76}, 354 (2016).

\bibitem{Chwastowski:2016zzl}
J. J. Chwastowski, S. Czekierda, R. Staszewski, M. Trzebi{\'n}ski, \textit{Eur. Phys. J. C} {\bfseries 77}, 216 (2017).

\bibitem{Adriani:2022csq}
O. Adriani \textit{et al.}, arXiv:2203.15416 [hep-ex].

\bibitem{LHCf:2017fnw}
LHCf Collaboration, O. Adriani \textit{et al.}, 
\textit{Phys. Lett. B} {\bfseries 780}, 233 (2018).

\bibitem{Tiberio:2023xgj}
LHCf Collaboration, A. Tiberio \textit{et al.}, 
\textit{PoS} ICRC2023 (2023) 444.

\bibitem{Lebiedowicz:2021byo}
P. Lebiedowicz, O. Nachtmann, A. Szczurek,
\textit{Phys. Rev. D} {\bfseries 105}, 014022 (2022).

\bibitem{Lebiedowicz:2013xlb}
P. Lebiedowicz and A. Szczurek,
\textit{Phys. Rev. D} {\bfseries 87}, 114013 (2013).

\bibitem{Ewerz:2013kda}
C. Ewerz, M. Maniatis, and O. Nachtmann,
\textit{Annals Phys.} {\bfseries 342}, 31 (2014).

\bibitem{Ewerz:2016onn}
C. Ewerz, P. Lebiedowicz, O. Nachtmann, and A. Szczurek,
\textit{Phys. Lett. B} {\bfseries 763}, 382 (2016).

\bibitem{Lebiedowicz:2013vya}
P. Lebiedowicz and A. Szczurek,
\textit{Phys. Rev. D} {\bfseries 87}, 074037 (2013).

\bibitem{Lebiedowicz:2016ioh}
P. Lebiedowicz, O. Nachtmann, A. Szczurek,
\textit{Phys. Rev. D} {\bfseries 93}, 054015 (2016).

\bibitem{Lebiedowicz:2023mhe}
P. Lebiedowicz, O. Nachtmann, A. Szczurek,
\textit{Phys. Rev. D} {\bfseries 107}, 074014 (2023).

\bibitem{Lebiedowicz:2023ell}
P. Lebiedowicz, O. Nachtmann, A. Szczurek,
arXiv:2307.13291 [hep-ph], accepted for publication in Phys. Rev. D.

\bibitem{Foroughi-Abari:2021zbm}
S. Foroughi-Abari and A. Ritz,
\textit{Phys. Rev. D} {\bfseries 105}, 095045 (2022).

\bibitem{Shin:2021bvz}
C. S. Shin and S. Yun,
\textit{JHEP} {\bfseries 02}, 133 (2022).

\bibitem{Gorbunov:2023jnx}
D. Gorbunov and E. Kriukova,
\textit{JHEP} {\bfseries 01}, 058 (2024).

\bibitem{FASER:2023tle}
FASER Collaboration, H. Abreu \textit{et al.}, 
\textit{Phys. Lett. B} {\bfseries 848}, 138378 (2024).

\bibitem{ALICE:2022wwr}
ALICE Collaboration, arXiv: 2211.02491 [physics.ins-det].

\end{thebibliography}
\end{document}